# Virtually Extended Coworking Spaces? – The Reinforcement of Social Proximity, Motivation and Knowledge Sharing Through ICT

## Completed research paper

**Lennart Hofeditz**
Department of Computer Science and Applied Cognitive Science
University of Duisburg-Essen
Duisburg, Germany
Email: lennart.hofeditz@uni-due.de

**Milad Mirbabaie**
Faculty of Business Studies and Economics
University of Bremen
Bremen, Germany
Email: milad.mirbabaie@uni-bremen.de

**Stefan Stieglitz**
Department of Computer Science and Applied Cognitive Science
University of Duisburg-Essen
Duisburg, Germany
Email: stefan.stieglitz@uni-due.de

## Abstract

Coworking is characterized by different people sharing a workspace to benefit from the inspiring working atmosphere. Even before Covid-19, many positive effects and dynamics were not fully exploited by their users. One reason is a lack of trust among the users that leads to social isolation, although a coworking space should increase knowledge and idea exchange. As most people in coworking spaces use information and communication technologies (ICT) for their collaboration with their clients or employers, we examined if and how ICT can be used to support the positive effects and dynamics of coworking spaces. For this, we conducted eight interviews with freelancers and entrepreneurs who have already worked in coworking spaces in order to identify requirements for a complementary virtual coworking platform. We found that social proximity, motivation and knowledge sharing could be increased by such a platform. Based on the process virtualization theory, we derived six design principles.

# 1 Introduction

Information and communication technologies (ICTs) and the digitization of organizations allow people to work from remote locations while traveling (Kong et al. 2019). Even though this has some advantages, it can also result in social isolation (von Zumbusch and Lalicic 2020). As many freelancers and people who are self-employed mostly work alone, many of them use coworking spaces to counteract social isolation and to increase their productivity (Bueno, Rodríguez-Baltanás, and M Dolores Gallego 2018). Coworking spaces are defined as "dynamic, inspiring and low-cost workplaces where people (from different business backgrounds) can interact, share knowledge and co-create" (Weijs-Perrée et al. 2019, p. 534). Alongside the possibility to share knowledge with others, individuals can build or expand their social networks in such work environments (Weijs-Perrée et al. 2019). Further advantages are positive effects on creativity, productivity and social learning (Bilandzic and Foth 2013; Blagoev et al. 2019; Brown 2017). However, the worldwide Covid-19 pandemic demonstrates that it is not always possible for face-to-face communication (Morrow 2020) in coworking spaces to take place. Even before Covid-19, many people were not aware of all advantages and positive effects of coworking spaces (Pollak and Anderst-kotsis 2017). Furthermore, there often is a lack of mutual trust to exchange ideas (Egea 2017), as it is often not clear whether other people in coworking spaces are interested in social and professional exchange or not.

For communicating with clients, freelancers and self-employed people mostly use ICTs such as Google Drive, MS Teams or Slack (Van Ostrand et al. 2016). Moreover, there is a wide range of studies that examined the collaboration and communication of distributed project teams (Andres et al. 2009; Hassell and Cotton 2017; Massey et al. 2014). However, platforms that support the exchange of users of coworking spaces rarely exist so far. Platforms such as sococo already try to establish as a coworking online platform (*Sococo - Online Workplace for Distributed Teams.* 2019). However, existing platforms are not empirically grounded or examined and rather provide the networking of freelancers beyond national borders than supporting people within coworking spaces. The question arises how ICTs can be used to decrease isolation of self-employed people and freelancers and to what extent we can learn from the existing information systems (IS) literature. Hence, we derived the following research question:

*RQ1: How can ICT be used to improve coworking spaces?*

Furthermore, there is a wide range of positive effects of coworking spaces, but not all of them are exploited by users (Egea 2017). Although ICT has been shown to improve collaboration and communication, it is hard to determine precisely which positive aspects of coworking spaces can be enhanced by a virtual platform. That is why we derived the following second research question:

*RQ2: Which positive aspects of coworking spaces can be improved by a complementary virtual platform?*

To answer these research questions, we not only considered literature on ICT, but also interviewed eight users of coworking spaces and identified requirements for a supportive virtual platform that could reinforce positive effects such as an increased level of productivity, creativity, networking, collaboration, knowledge sharing and motivation. To structure our interviews we designed our interview guide according to the process virtualization theory (PVT) of Overby (2011). Therefore, the aim of this paper is to contribute to the research stream of digital work and virtual collaboration in IS. We examined coworking spaces from an information systems (IS) perspective and identified six design principles for a complementary virtual platform based on the requirements of our interviewees and corresponding IS literature.

# 2 Background

## 2.1 Coworking Spaces

In 2017, the number of coworking spaces rose to 15,500 places worldwide (Vidaillet and Bousalham 2018). This strong growth is accompanied by an even higher future forecast for the number of coworking spaces (Andrade et al. 2013). Users of a coworking space can rent a workplace for a rather small amount of money and benefit from the inspiring working atmosphere which is generated by diverse people sharing a workspace (Weijs-Perrée et al. 2019). Overall, such working environments enable the collective use of infrastructure and resources such as desks, meeting rooms and WI-FI connections (Bouncken and Reuschl 2018). Additionally, employees can share knowledge, brainstorm together and provide emotional support (Spinuzzi et al. 2019). Coworking space activities are meant to contribute to the life quality of employees, not only in terms of work, but also in relation to private life. They were developed to avoid the feeling of loneliness when working from home and to enable people to separate their private and working lives (Weijs-Perrée et al. 2019). However, the expectation of reduced loneliness is not always fulfilled. Most of the users of coworking spaces are self-employed workers, small firms such as startups or micro-businesses and students (Weijs-Perrée et al. 2019). Bouncken and Reuschl (2018) stressed that increased social interactions are one of the main reasons for working in coworking spaces but in fact working alone together has many more advantages. New skills can be acquired in a coworking environment, because coworking spaces are characterized by members who all have different knowledge and experiences, which leads to a huge variety and opportunities to learn new things (Bueno, Rodríguez-Baltanás, and M. Dolores Gallego 2018). According to Bouncken and Reuschl (2018), the co-presence of people with different or similar attitudes and characteristics can result in increased satisfaction and as a result economic benefit. Thus, the presence of other people provides positive working experiences and outcomes as positive emotions and experiences have a positive impact on job satisfaction. Furthermore, coworking environments and social interactions can have a positive influence on productivity (Bueno, Rodríguez-Baltanás, and M. Dolores Gallego 2018). Creativity can be increased by sharing knowledge and inspiration through interaction with other coworkers (Bouncken and Reuschl 2018). In addition, it could be shown that coworking can even increase employees' overall performance and self-efficacy (Bouncken and Reuschl 2018). All in all, coworking can improve the performance of independent workers (Bouncken and Reuschl 2018; Garrett et al. 2017; Weijs-Perrée et al. 2019).

However, there are also some negative aspects. Although the ability to communicate with other users is an advantage, the predominant and often too much noise in coworking spaces can be criticized (Bouncken 2018). Literature on coworking spaces indicated that they would have many positive effects on their users, but it is unclear whether they are perceived and valid in practice. Furthermore, interactions such as brainstorming or knowledge exchange require a certain amount of extraversion and self-confidence. Due to a high level of anonymity, it is also questionable how well networking in coworking spaces really works. The contact restrictions imposed by Covid-19 could also contribute to the lack of a lively exchange of ideas in coworking spaces.

For collaboration, it could already be shown that the correct use of ICT can help to mitigate and overcome similar issues (Egea 2017). The question arises how ICT can be used to virtualize the positive aspects of coworking to provide a complementary digital workplace and how IS research on virtual collaboration can contribute to design a digital platform that supports knowledge sharing in coworking spaces.

## 2.2  Collaboration vs. Coworking at Coworking Spaces

Coworking spaces are sparsely examined in IS research to this time (Josef 2018; Kong et al. 2019; Schlagwein 2018). They are usually only discussed in passing within the context of digital nomads, remote work or flexible work concepts (Schlagwein and Jarrahi 2020). To shape how ICT can be used to improve coworking spaces, it is important to understand the activities of people working at such spaces. As the general activity practiced at coworking spaces, coworking can be described as "emergent collaborative activity", which emphasizes that coworking and collaboration are interrelated (Spinuzzi et al. 2019). Collaboration therefore is a more specific activity at coworking spaces that could be positively influenced by a complementary platform. There are some key elements which highlight the differences between coworking and collaboration. Referring to Mattessich and Monsey (1993) collaboration is characterized by a persuasive relationship. Coworking is more dynamic, because people could change the coworking spaces every time. This prevents permanent working groups which are not necessary in coworking spaces. Spinuzzi (2019) described coworking as "Working alone together" (p. 399). In contrast, collaboration requires cohesion and arrangement (Reeves et al. 2018). A common feature of coworking and collaboration is that in both people from different disciplines and interests come together (Reeves et al. 2010; Weijs-Perrée et al. 2019). One the one hand, collaboration is therefore defined as a process where a group of people work together cooperatively to complete a problem-solving task (Alavi et al. 1995). Coworking, on the other hand, may involve collaboration, but usually involves the aspect of 'working alone together', with each individual pursuing his or her own goals (Spinuzzi 2012). Whereas collaboration can be supported by ICT such as enterprise social networking platforms (Laumer et al. 2017) or enterprise collaboration systems (Nitschke et al. 2019), coworking has not yet been considered as a process that can be supported by ICT.

### 2.3  Process Virtualization Theory

For a theory-guided virtualization of physical work processes a process virtualization theory (PVT) can be used (Overby 2008). The PVT describes how processes can be transformed virtually. For coworking spaces, the theory provides a guided approach to consider processes and dynamics in order to test their virtualizability. By doing so, the theory tries to explain to what extent factors of a process or a sub-task can be performed virtual and automated (Overby 2011). To measure this virtualizability there are two approaches used in research. On the one hand it can be measured how often a process is transferred into a virtual environment to derive a specific level of need. On the other hand, the quality of a process in a physical and a virtual environment can be compared. If the quality is much higher in the virtual version of the process the virtualizability is very high. The theory not only considers the states "physical process" and "virtual process" but also the characteristics of the process and the characteristics of the virtualization mechanism (Overby 2011). In this study we tested whether the theory is applicable to identify requirements for a complementary platform for users of coworking spaces. The PVT consists of four main constructs that describe the characteristics of processes: a) the *sensory requirements* which refer to the importance of the main sensory impressions for a process, b) the *relationship requirements* which include the role of all types of social interactions for a process, c) the *synchronism requirements* which refer to place and time of a process and to whether the respective communication needs to have a direct response or not and d) the *identification and control requirements* which include both the identification of all participants in the process and the control over their behavior. The PVT is visualized in Figure 1.

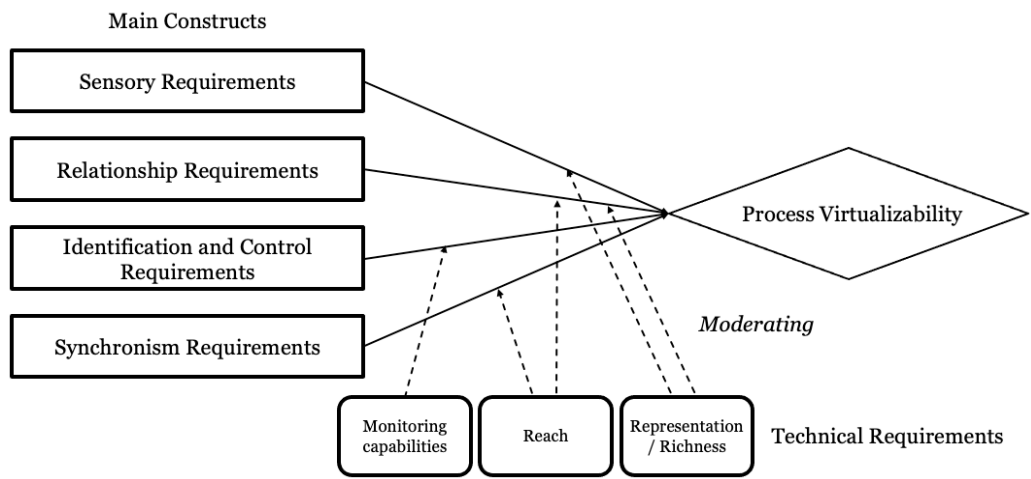

*Figure 1. Process Virtualization Theory (Overby 2011)*

These constructs of process virtualizability can be affected by IT-related mechanisms especially when the considered case include technology. Grounded on the PVT, we derived our research design. However, we were aware of some limitations of the PVT, as it was designed to show the virtualizability of one physical process and not a set of processes such as the dynamics at coworking spaces. In our research, we were therefore open to additional aspects that might not be covered by the PVT.

## 3  Research Design

Since empirical research on ICT use in coworking spaces is very limited, detailed experiences and field reports are needed, in order to make specific statements to this topic. In order to identify which positive effects of coworking spaces can be supported by a virtual platform and how, we decided to conduct semi-structured interviews with different groups of people working at coworking spaces. We recruited employees, freelancers, entrepreneurs and students who continuously used coworking spaces, who were familiar with remote full- or part-time work, and who also were experienced in using ICT at the workplace. We chose these four user groups because we found those in the literature to be the main user groups of coworking spaces (Bouncken 2018; Bouncken and Reuschl 2018; Spinuzzi 2012). We recruited both full-time and part-time interviewees working in coworking spaces in order to be able to depict a composition of people in coworking spaces which is as close to actual situations as possible and to be able to identify design principles which are useful for a broad range of people. In addition, we discovered in our search that these groups often overlap. For example, students can work as freelancers or with start-ups at the same time, and companies often employ freelancers for a short or a longer time period. After eight extensive interviews with recruited participants we had the impression that we covered the main types of coworking space users. Another reason besides the perceived saturation was that we analyzed additional findings from previous literature as a second source of data to establish the design principles for a complementary virtual coworking platform.

For the interviews, we used a guideline which consisted of open questions about what they think about the virtualization of specific processes at coworking spaces and how a complementary platform needs to be designed to be useful for them. To structure the guideline, we asked for *sensory requirements*, *relationship requirements*, *identification and control requirements* and *synchronism requirements* according to the PVT (Overby 2011). In addition, we guided our interviews by following the constructs of the collaboration virtualization theory which was used to virtualize collaboration processes (Fan et al. 2013). As an example, referring to *urgency* and *complexity* of the category *task,* the interviewees had to picture what communication tools they would like to use in a particularly urgent or complex task (sample question*: 'If a task is particularly urgent, what communication tools would you prefer to use?'*). We also enquired the influence of the presence of others on their own working performance (sample question: *'How should other coworking space users be represented on a virtual platform?'*). The questions were primarily related to previous experiences of relationships in physical coworking spaces. However, we asked to what extent the recipients want to communicate with others via an online platform (sample question*: 'How do you imagine the general network among online coworkers?'*). In addition to our open questions and sub-questions, we presented three mock-ups in the interviews. They were provided as an anchor to imagine a complementary coworking platform and to give possible thought-provoking impulses regarding the preferred design and functions. We showed the mockups after the interviewees had already been asked about a possible coworking platform features, in order to prevent them from biasing the answers. We created two mock-ups ourselves including features of different collaboration tools and a possible representation of an office from different perspectives (Slack, Bitrix24, Roomsketcher). The second one was a modified screenshot of an existing online coworking platform, that tries to offer an alternative to physical coworking spaces (*Sococo - Online Workplace for Distributed Teams.* 2019). Figure 2 shows the self-created mock-up of a coworking platform. The second mock-up differs only in perspective.

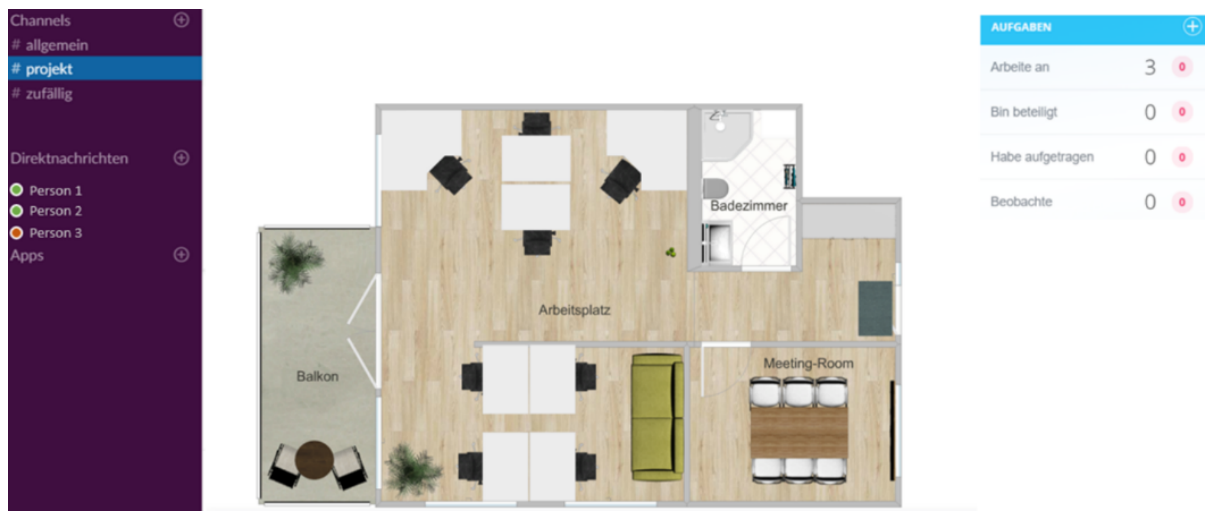

*Figure 2: Mock-up of a platform that could support people in coworking spaces which was shown in the interviews.*

We identified the participants through internet research and existing contacts (peers, colleagues and project partners) and conducted the interviews between May and June in 2019. The duration of the interviews varied between 40 and 70 minutes with an average duration of 49 minutes.  As there are different groups of people working in coworking spaces such as employees, students, entrepreneurs and freelancers (Weijs-Perrée et al. 2019), we interviewed people representing one or several of these groups (Table 1.).

*Table 1. Description of the interviewed coworking space users*

| Synonym | Student | Freelancer | Entrepreneur | Employee | ICT experience | Age | Gender |
|---|---|---|---|---|---|---|---|
| S1 | X | X | | | X | 26 | female |
| S2 | X | | | X | X | 22 | female |
| S3 | X | X | | | X | 28 | female |
| F1 | | X | | | X | 31 | male |
| F2 | | X | | | X | 27 | female |
| Em | | | | X | X | 32 | male |
| En1 | | | X | | X | 30 | male |
| En2 | | | X | | X | 31 | male |

We transcribed the interviews literally and analyzed them using MAXQDA, as it is a software especially designed for computer-assisted qualitative and mixed methods to analyze text and multimedia data (Saillard 2011). For analyzing the data, we applied a qualitative content analysis according to Mayring (2014) and followed a deductive approach for category formation. Therefore, we created the categories according to the constructs of the PVT and some aspects of the collaboration virtualization theory. In the first step, we paraphrased corresponding text passages. For the coding, three experienced coders discussed every paraphrase against the background of the PVT categories and excluded paraphrases which did not match. If a phrase was considered relevant but did not match with the initial categories, an additional category was derived. After the coding, we applied a generalization step according to Mayring (2014) and interpreted our results. We then developed design principles based on current literature findings and the findings from the expert interviews which were conducted and followed the guidelines of Kruse et al. (2016).

# 4   Results & Discussion

## 4.1   Derivation of Requirements and Design Principles

**Sensory requirements:** Our interviewees highlighted that one of the greatest advantages of physical coworking spaces is the emotional support and the perceived motivation which could also be increased by presenting logged-in users on the platform (F1-2, Em, En2). As an example, S1 stated:" *I believe that it is a kind of help to motivate oneself to work. Somehow to sit down in a nice environment, in a nice working atmosphere. To meet other people maybe, so mostly positive people, who also, mostly I noticed, who either learn or do creative things*". Weiss and Cropanzano (Weiss and Cropanzano 1996) described that cognitive and emotional processes have an influence on job satisfaction which can result in an increase of productivity. Emotional support during work is very important and should always be sought. A virtual coworking platform would provide one opportunity to disseminate this effect beyond the borders of physical coworking spaces to people who would like to work from home, from a café, from a library, or from other open spaces. To represent the emotional support in a virtual environment, auditive and visual applause was mentioned in our interviews (S3). This could also be mixed with other technologies providing emotional support such as automated virtual assistants (Letho et al. 2011). In this way, coworkers are motivated even without the physical presence of others. If, for example, a project has been successfully completed or if things are not going well, other coworkers can virtually express their happiness or empathy:

*DP1: Provide a complementary virtual coworking platform with the possibility to get emotional encouragement in order to ensure that users can also get virtual auditory and visual feedback when they are in social or physical absence of others.*

**Relationship requirements:** In collaboration, trust within a group is seen as crucial to team success, especially in terms of knowledge sharing and virtual collaboration (Fan et al. 2013). This raises the question of the extent to which relationships are necessary in coworking and virtual coworking. Fan et al. (Fan et al. 2013) concluded that strong relationships and thus collaboration are easier to virtualize. The relationship between users of coworking spaces was described as familiar and friendly, nevertheless professional and collegial in our interviews (F1, En2). In some cases, personal contacts and friendships were formed (F1-2, Em, En2). F1 stated: *"So I met a few coworkers with similar age privately. Simply because there were a few overlapping interests, a few friendships were formed. But then that remained rather private"*. The interviewees emphasized that the contact is usually not as intense as it is supposed to (S1, S3, F1, Em). They pointed out the disadvantage that everyone in the coworking space works on their individual tasks and that there is little exchange. Interviewee F2 stressed that coworking does not create synergies in terms of collaboration. This contradicts the definition of coworking, which assumes that people interact and co-create (Weijs-Perrée et al. 2019). However, all interviewees said that having productive people around increases their own motivation. Thus, it seems valuable to see who is active and at work. The interviewees suggested to implement icons and self-designed avatars or simple profile pictures for a better overview. In addition, a status consisting of a few words could indicate whether the person is currently available or busy (F1-2, Em, En2). A symbol was wanted, which shows the availability status of each person (S1-2, En1). This is consistent with statements made by Dourish and Bellotti (Dourish and Bellotti 1992) which stressed that awareness, such as knowing what others are doing, is important. This finally results in our following design principle:

*DP2: Ensure that user motivation in the social or physical absence of others is supported by a complementary virtual coworking platform by transparently displaying logged-in users through profile pictures and an availability indicator.*

According to the interviews, users of coworking spaces are placed by their competences and areas of responsibility. Meeting people with similar attitudes has a positive influence on satisfaction and thus on work performance (Bouncken and Reuschl 2018). The interviewees were already familiar with the function of channels from collaboration tools (such as *slack*), which facilitates their use and integration. In addition to the functionalities already available in collaboration tools, three new ones were named that would complement coworking platforms. A browsing mode would take you into a virtual coworking world where things and people can be discovered (S3). Invisibility mode would mean not only making your status 'busy' but disappearing completely from the platform for a certain period (S1): *"Maybe also such a function that you can make yourself invisible, that nobody can contact you or see that you are there. I think that would be quite good"*. Furthermore, the electronic brainstorming or virtual idea exchange area describes an area in which one presents ideas on the platform and gets feedback from other coworkers (En2). In general, individualization should be offered to a certain extent, since this heightens the degree of involvement, which in turn could have a positive effect on the intention to use a coworking platform (En2). Private and professional exchange are to be separated (F1, En2). It was also explicitly stated that small talk – so-called kitchen conversations – should arise just as in the physical coworking space (S2-03, En1). Interviewee F1 believed that there is a greater chance and less restraint in contacting others online than in coworking spaces. This results in our following design principle:

*DP3: Provide a complementary virtual coworking platform with the ability to exchange information in different channels in order to ensure that users who have the same interests can find each other easily in order to increase the knowledge exchange.*

**Synchronism Requirements:** Many interviewees stated that in the case of urgent tasks they would interact via synchronous communication that is most likely face-to-face communication (F1, Em, En1, En2). Asynchronous communication, such as email, would be disadvantageous in this situation (S3, Em). Faster systems, such as messengers are more important here (S1, S3, Em, En2). These especially include phone calls and video conferences to be able to express something more directly, avoid misunderstandings and to act more efficiently (S1-3, F1-2, En2). In addition, it is always important to document the discussed content in the form of a written note, such as through email (Em). Furthermore, a desktop sharing tool was desired in order to be able to clarify contents in combination with a conversation (S1-2, F1). This is supported by researchers who argued that there are less delays when everyone works at the same place (Baltes et al. 2002; Fan et al. 2013). In practice there are both synchronous and asynchronous communication features (*Sococo - Online Workplace for Distributed Teams.* 2019). According to the process virtualization theory the required synchronicity contributes to lower virtualizability (Overby 2008). In order to avoid this problem, we have identified a suggested solution in the interviews. A ticket system could help to highlight particularly urgent tasks and ensure faster communication. In addition, mobile phone notifications facilitate the speed of answers. If respondents work on a complex task, they would also prefer synchronous communication. A phone call or a video conference (En2) can be used to solve problems and create a better understanding. To provide an example, En2 stated: *"I believe that coordination can be achieved faster with video telephony or instant messaging, for example. Or with a short telephone conference"*. It makes sense to share the screen in order to be able to describe complex issues more easily (S1). It was additionally noted that asynchronous communication tools, such as messengers, a forum, organization tools and emails should be used in such situations (S2-3, En1). According to Kirkman and Mathieu (2005) complex tasks primarily require technologies that provide rich information and synchronous communication channels. However, our results showed that it is not sufficient to communicate only synchronously. This could be due to the fact that especially when working alone together (as it is the case in coworking spaces) it is important to make communication available for a longer period of time. This is also reflected in the interviewees' statements, as they would use both synchronous and asynchronous communication channels. These results leaded us to develop the following design principle:

*DP4: Provide a complementary virtual coworking platform with different types of computer-mediated communication in order to ensure that users can perform synchronous or asynchronous communication to complete urgent and complex tasks.*

**Identification and Control Requirements:** According to Lee (Lee 2009) internet-based communication can lead to the unauthorized transfer of data. The fear that unauthorized persons could get access to personal and sensitive data, was also expressed in our interviews. But in general, our interviewees had no concerns regarding the privacy of personal data on a complementary virtual coworking platform, as they perceive basic trust in such platforms. However, that does not mean that personal data privacy is no important factor. When it comes to job-related data, more problems and concerns are raised by the interviewees. Even in physical coworking, it can occur that co-users get to know work-related information that should be kept protected. This problem is anticipated when working on digital workplaces. For example, there could be damage to their businesses if a competitor sees sensitive data or when data is passed on to third parties or is lost if the server leaks data (S2-3, Em, En1-2). As an example, S3 said: *"Basically, I would say that there are concerns about whether the data is read out for the purposes of third parties"*. Therefore, it should be ensured that sensitive and private files are protected on a coworking platform. Kamat et al. (2003) also claimed that access rights are needed to protect the data online. Even in coworking spaces strangers can look into work-related documents (Bouncken and Reuschl 2018). Thus, access rights are necessary to protect the professional data. This aspect is already highlighted in literature, since it is a well-established way of protecting data (Fan et al. 2013; Kamat et al. 2003). From these data protection issues, we formulated the following design principle in order to determine who can view and edit which information and data on a coworking platform:

*DP5: Ensure that a complementary virtual coworking platform is aligned with the security of confidential data in order to ensure that users can control the way their job-related files are handled in terms of access rights and permissions.*

**Organizational requirements:** The extent to which the technology enables the virtualization of collaboration processes depends on the wealth of information and the type and number of communication channels (Fan et al. 2013). According to the PVT, a higher information technology capacity will make collaboration more virtualizable (Overby 2011). Our interviews revealed that standard tools such as text, voice or video chat should be included. However, organizational and planning tools such as project management tools play an important role (S2-3, En1). Interviewees S2 and F1 also believed that a common digital pinboard would be useful: *"Tiles with different buckets, which can be layered up and down according to priority, where comments are attached, where the task is described, where files may be attached and where it says who is going to do it and until when. Basically, a shared pinboard"* (S2). Thereby, information and ideas can be shared and opinions can be exchanged. Research showed that a digital display with information about present people leads to more social interaction and knowledge sharing (Parrino 2015). Other researchers already argued that online coworking could be managed by a coordinator, which can be a user or an artificial intelligence (Andrade et al. 2013). The use of the tool also depends on whether the platform is used only at a physical coworking space or also at other places where people still want to feel part of the coworker community. In our interviews we discovered that only a few of the interviewees participate at events in coworking spaces so far (S2, F1-O2). This could be due to poor organization. Participation may increase through appropriate organizational tools such as embedded calendar functionalities or virtual assistant support (Diederich et al. 2019). Therefore, our sixth design principle is the following:

*DP6: Provide a complementary virtual coworking platform with organizational tools to ensure that all members of the coworking platform can clarify organizational issues even if they are not at the physical coworking space.*

## 4.2 Implications

In general, it can be stated that some positive effects and mechanisms of physical coworking could be supported by a complementary virtual coworking platform. Especially motivation, knowledge sharing and social proximity could be increased according to our interviewees. The PVT was well suited as a framework to identify design principles for a complementary virtual platform. The constructs of the theory allowed us to cover almost all relevant aspects in order to test the processes and dynamics in coworking spaces for virtualization. However, the PVT was designed to test the virtualizability of physical processes. A complementary virtual coworking platform could deliver further advantages that are not found in physical coworking. For these aspects not covered by the PVT, we added the requirement category of organizational requirements which resulted in DP6. Although the technical requirements of PVT (monitoring capabilities, reach and representation) also include organizational aspects, our interviews showed that the creation of organizational requirements is highly relevant. This aspect could also be highly relevant for other complex processes or process chains that needs to be virtualized. We have therefore added organizational requirements to the PVT model (Figure 3).

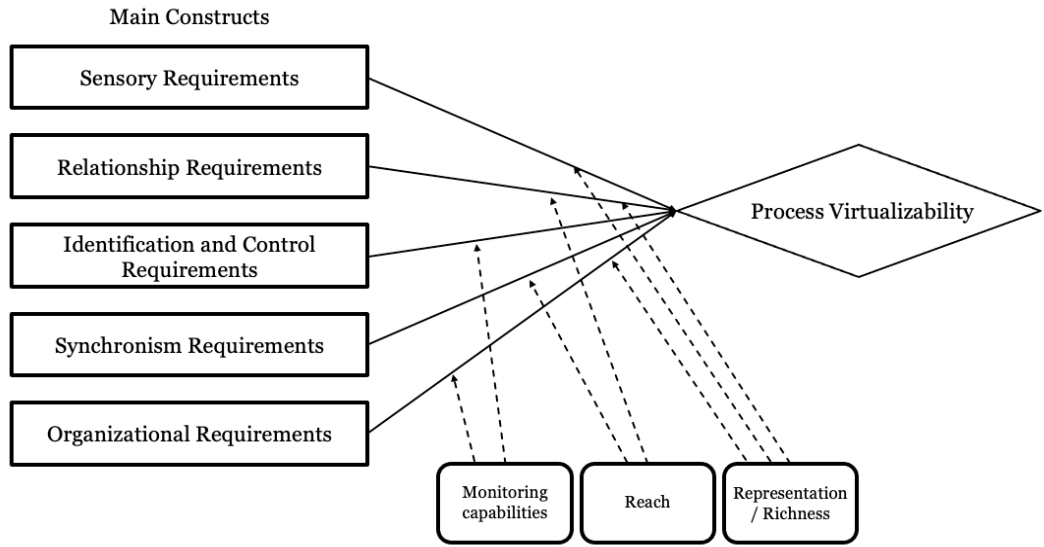

*Figure 3. Modified PVT showing the virtualizability of dynamics at coworking spaces*

The organizational requirements represent all organizational aspects of a process, mechanism or a set of processes such as organizing tasks or meetings that are amenable or resistant to being conducted virtually. These requirements are also influenced by technical aspects such as monitoring capabilities and representation and richness.

Moreover, almost no process today is completely physical or completely virtual. Most processes consist of hybrids of physical interaction and virtual operations. PVT assumes completely physical processes for virtualization, but even the processes in coworking spaces contain some virtual elements that can be linked to. Further research should examine the virtualizability of these mixed processes in order to identify more requirements that makes a process even more effective when it is performed virtually.

### 4.3 Limitations & Further Research

The present paper also contains limitations that can be considered with regard to future research. It would be valuable to interview more people with experience in virtual collaboration, in order to identify more concrete requirements. Moreover, this study is limited to Germany, but it would also be interesting to discover intercultural and country-specific differences. There is also a need to investigate how to ensure that current problems such as technostress can be avoided by using a coworking platform (Ayyagari et al. 2017). Furthermore, this study we focused on an ICT solution. Future research should examine other aspects that could improve the work at coworking spaces such as emotional factors.

Finally, it should be further investigated whether there are differences in the functionalities of a virtual coworking platform when it is used at home or only in addition to a physical coworking space, and which opportunities both options can offer. Furthermore, the developed design principles should be tested to examine the practicability and the need for the actual use of the system. Future research should first focus on verifying the conceptual model of coworking virtual reinforcement theory which we provided in this study.

## 5 Conclusion

In this study, we found that positive effects and dynamics of coworking spaces can be supported by complementary virtual coworking platforms. Especially motivation, knowledge sharing and social proximity of a coworking space could be increased by such a platform. We showed how ICT, which is already used to communicate with clients, can also be used to increase motivation, social exchange and creativity. In addition, we showed that the PVT provides a first guided approach to identify requirements for a complementary virtual coworking platform. Furthermore, we identified organizational requirements as an important factor that needs to be considered when virtualizing the dynamics and set of processes at coworking spaces. We derived six design principles from our interview findings and corresponding literature which should be taken into account when designing such a platform. The platform should serve as a supplement to a physical coworking space and thus enable participants, who are not always able to physically interact with others, to be a part of the community and to benefit from the advantages such as knowledge and motivation exchange. The exploration of such a platform from an IS perspective is important as it is primarily focusing on processes and effects of coworking spaces and not on the physical coworking space itself.

# 6 References

Alavi, M., Wheeler, B. C., and Valacich, J. D. 1995. "Using IT to Reeingineer Business Education: An Exploratory Investigation of Collaborative Telelearning," *Mis Quarterly*.

Andrade, J., Ares, J., Suárez, S., and Giret, A. 2013. "Implementation Challenges for Supporting Coworking Virtual Enterprises," *Proceedings - 2013 IEEE 10th International Conference on e-Business Engineering, ICEBE 2013*, pp. 9–16. (https://doi.org/10.1109/ICEBE.2013.2).

Andres, H. P., Shipps, B. P., and Carolina, N. 2009. *Team Learning in Technology-Mediated Distributed Teams*, (21:2), pp. 213–222.

Ayyagari, Grover, and Purvis. 2017. "Technostress: Technological Antecedents and Implications," *MIS Quarterly* (35:4), p. 831. (https://doi.org/10.2307/41409963).

Baltes, B. B., Dickson, M. W., Sherman, M. P., Bauer, C. C., and LaGanke, J. S. 2002. "Computer-Mediated Communication and Group Decision Making: A Meta-Analysis," *Organizational Behavior and Human Decision Processes* (87:1), pp. 156–179. (https://doi.org/10.1006/obhd.2001.2961).

Bilandzic, M., and Foth, M. 2013. "Libraries as Coworking Spaces: Understanding User Motivations and Perceived Barriers to Social Learning," *Library Hi Tech* (31:2), pp. 254–273. (https://doi.org/10.1108/07378831311329040).

Blagoev, B., Costas, J., and Kärreman, D. 2019. "'We Are All Herd Animals': Community and Organizationality in Coworking Spaces," *Organization*, pp. 1–23. (https://doi.org/10.1177/1350508418821008).

Bouncken, R. B. 2018. "University Coworking-Spaces: Mechanisms, Examples, and Suggestions for Entrepreneurial Universities," *International Journal of Technology Management*. (https://doi.org/10.1504/ijtm.2018.10012930).

Bouncken, R. B., and Reuschl, A. J. 2018. "Coworking-Spaces : How a Phenomenon of the Sharing and for Entrepreneurship," *Review of Managerial Science* (12:1), Springer Berlin Heidelberg, pp. 317–334. (https://doi.org/10.1007/s11846-016-0215-y).

Brown, J. 2017. "Curating the 'Third Place'? Coworking and the Mediation of Creativity," *Geoforum* (82:April), Elsevier, pp. 112–126. (https://doi.org/10.1016/j.geoforum.2017.04.006).

Bueno, S., Rodríguez-Baltanás, G., and Gallego, M Dolores. 2018. "Coworking Spaces: A New Way of Achieving Productivity," *Journal of Facilities Management* (16:4), pp. 452–466. (https://doi.org/10.1108/JFM-01-2018-0006).

Diederich, S., Brendel, A. B., and Kolbe, L. M. 2019. "On Conversational Agents in Information Systems Research: Analyzing the Past to Guide Future Work," *Proceedings of the International Conference on Wirtschaftsinformatik*.

Dourish, P., and Bellotti, V. 1992. "Awareness and Coordination in Shared Workspaces," *CSCW '92 Proceedings of the 1992 ACM Conference on Computer-Supported Cooperative Work*, pp. 107–114.

Egea, K. 2017. "Relationship Building in Virtual Teams: An Academic Case Study," *Proceedings of the 2006 InSITE Conference*. (https://doi.org/10.28945/3046).

Fan, S., Choon, L. S., and J. Leon, Z. 2013. "Towards Collaboration Virtualization Theory," *PACIS 2012 Proceedings*, pp. 1–10. (https://doi.org/10.1002/pbc.ABSTRACT).

Garrett, L. E., Spreitzer, G. M., and Bacevice, P. A. 2017. "Co-Constructing a Sense of Community at Work: The Emergence of Community in Coworking Spaces," *Organization Studies* (38:6), pp. 821–842. (https://doi.org/10.1177/0170840616685354).

Hassell, M. D., and Cotton, J. L. 2017. "Some Things Are Better Left Unseen: {Toward} More Effective Communication and Team Performance in Video-Mediated Interactions," *Computers in Human Behavior* (73), pp. 200–208. (https://doi.org/10/gbk9cw).

Josef, B. 2018. *Coworking from the Company's Perspective - Serendipity-Biotope or Getaway-Spot?*, pp. 265–278. (https://doi.org/10.18690/978-961-286-043-1.19).

Kamat, N., Vishwanathan, S., Prabhakar, B. S., Goradia, T., and Saran, A. 2003. *Dynamic Rules-Based Secure Data Access System for Business Computer Platforms*, (1:19).

Kirkman, B. L., and Mathieu, J. E. 2005. "The Dimensions and Antecedents of Team Virtuality," *Journal of Management* (31:5), pp. 700–718. (https://doi.org/10.1177/0149206305279113).

Kong, D., Schlagwein, D., and Cecez-Kecmanovic. 2019. "Issues in Digital Nomad – Corporate Work : An Institutional Theory Perspective," in *Proceedings of the 27th European Conference on Information*, Stockholm, pp. 1–16.

Kruse, L. C., Seidel, S., and Purao, S. 2016. "Making Use of Design Principles," in *Lecture Notes in Computer Science*, Springer International Publishing, pp. 37–51. (https://doi.org/10.1007/978-3-319-39294-3).

Laumer, S., Shami, N. S., Muller, M., and Geyer, W. 2017. "The Challenge of Enterprise Social Networking (Non-)Use at Work: A Case Study of How to Positively Influence Employees' Enterprise Social Networking Acceptance," *Proceedings of the ACM Conference on Computer Supported Cooperative Work, CSCW*, pp. 978–994. (https://doi.org/10.1145/2998181.2998309).

Lee, S. J. 2009. "Online Communication and Adolescent Social Ties : Who Benefits More from Internet Use ? ∗," *Journal of Computer-Mediated Communication Online* (14), pp. 509–531. (https://doi.org/10.1111/j.1083-6101.2009.01451.x).

Letho, T., Oinas-Kukkonen, H., and Drozd, F. 2011. "FACTORS AFFECTING PERCEIVED PERSUASIVENESS OF A BEHAVIOR CHANGE SUPPORT SYSTEM," *Ejisdc* (45:6), pp. 1–14. (https://doi.org/10.1007/s40596-014-0205-9).

Massey, A. P., Montoya-weiss, M. M., Hung, Y., Massey, A. P., and Montoya-weiss, M. M. 2014. *Because Time Matters : Temporal Coordination in Global Virtual Project Teams Because Time Matters : Temporal Coordination in Global Virtual*, (1222). (https://doi.org/10.1080/07421222.2003.11045742).

Mattessich, P. W., and Monsey, B. R. 1993. *Collaboration: What Makes It Work - A Review of Research Literature on Factors Influencing Successful Collaboration*, Minnesota: Amherst H. Wilder Foundation.

Mayring, P. 2014. *Qualitative Content Analysis*.

Morrow, J. A. 2020. "Alone Together : Finding Solidarity in a Time of Social Distance," *Space and Culture* (23:3), pp. 315–319. (https://doi.org/10.1177/1206331220938643).

Nitschke, C. S., Williams, S. P., and Schubert, P. 2019. "A Multiorganisational Study of the Drivers and Barriers of Enterprise Collaboration Systems-Enabled Change," *International Conference on Wirtschaftsinformatik (WI)*, pp. 1612–1626.

Van Ostrand, A., Islands, C., Wolfe, S., Islands, C., Arredondo, A., Islands, C., Skinner, A. M., Islands, C., Visaiz, R., Islands, C., Jones, M., Islands, C., Jenkins, J. J., and Islands, C. 2016. *Creating Virtual Communities That Work : Best Practices for Users and Developers of E-Collaboration Software*, (12:4). (https://doi.org/10.4018/IJeC.2016100104).

Overby, E. 2008. "Information Technology Process Virtualization Theory and the Impact of Information Technology," *Organization Science* (19:2). (https://doi.org/10.1287/orsc.1070.0316).

Overby, E. 2011. "Process Virtualization Theory and the Impact of Information Technology.," *Academy of Management Proceedings* (2005:1), pp. G1–G6. (https://doi.org/10.5465/ambpp.2005.18781442).

Parrino, L. 2015. "Coworking: Assessing the Role of Proximity Knowledge Exchange," *Knowledge Management Research and Practice* (13), pp. 261–271.

Pollak, M., and Anderst-kotsis, G. 2017. "E-Mail Monitoring and Management with MS Social Bots," *IiWAS '17*, pp. 4–6. (https://doi.org/10.1145/3151759.3151799).

Reeves, S., Lewin, S., Espin, S., and Zwarenstein, M. 2010. *Interprofessional Teamwork for Health and Social Care*, Blackwell Publishing Ltd.

Reeves, S., Xyrichis, A., and Zwarenstein, M. 2018. "Teamwork, Collaboration, Coordination, and Networking: Why We Need to Distinguish between Different Types of Interprofessional Practice," *Journal of Interprofessional Care* (32:1), Taylor & Francis, pp. 1–3. (https://doi.org/10.1080/13561820.2017.1400150).

Saillard, E. K. 2011. "FORUM : QUALITATIVE SOCIAL RESEARCH Systematic Versus Interpretive Analysis with Two CAQDAS Packages : NVivo and MAXQDA," *Forum: Qualitative Social Research (QSR)* (12:1).

Schlagwein, D. 2018. "'Escaping the Rat Race': Justifications in Digital Nomadism," *Research-in-Progress Papers*.

Schlagwein, D., and Jarrahi, M. H. 2020. "The Mobilities of Digital Work : The Case of Digital Nomadism," in *Twenty-Eighth European Conference on Information Systems (ECIS2020)*.

*Sococo - Online Workplace for Distributed Teams*. 2019.

Spinuzzi, C. 2012. "Working Alone Together: Coworking as Emergent Collaborative Activity," *Journal of Business and Technical Communication* (26:4), pp. 399–441. (https://doi.org/10.1177/1050651912444070).

Spinuzzi, C., Bodrožić, Z., Scaratti, G., and Ivaldi, S. 2019. "'Coworking Is About Community': But What Is 'Community' in Coworking?," *Journal of Business and Technical Communication* (33:2), pp. 112–140. (https://doi.org/10.1177/1050651918816357).

Vidaillet, B., and Bousalham, Y. 2018. "Coworking Spaces as Places Where Economic Diversity Can Be Articulated: Towards a Theory of Syntopia," *Organization* (Ea 2354). (https://doi.org/10.1177/1350508418794003).

Weijs-Perrée, M., van de Koevering, J., Appel-Meulenbroek, R., and Arentze, T. 2019. "Analysing User Preferences for Co-Working Space Characteristics," *Building Research and Information* (47:5), Taylor & Francis, pp. 534–548. (https://doi.org/10.1080/09613218.2018.1463750).

Weiss, H. M., and Cropanzano, R. 1996. "Affective Events Theory: A Theoretical Discussion of the Structure, Causes and Consequences of Affective Experiences at Work," in *Research in Organizational Behavior: An Annual Series of Analytical Essays and Critical Reviews* (Vol. 18), B. M. Staw and L. L. Cummings (eds.), US: Elsevier Science/JAI Press, pp. 1–74.

von Zumbusch, J. S. H., and Lalicic, L. 2020. "The Role of Co-Living Spaces in Digital Nomads' Well-Being," *Information Technology & Tourism* (0123456789), Springer Berlin Heidelberg. (https://doi.org/10.1007/s40558-020-00182-2).